\documentclass[a4paper]{article}
\usepackage{indentfirst}
\usepackage{hyperref}
\usepackage{multirow}
\usepackage{latexsym,bm,amsmath,amssymb}
\usepackage{graphicx}
\usepackage{epsfig}
\usepackage{subfigure}
\usepackage{cite}
\usepackage{color}
\usepackage[top=1in, bottom=1in, left=1.2in, right=1.2in]{geometry}
\usepackage{appendix}
\usepackage{ulem}
\usepackage{epstopdf}

\newcommand{\zc}{Z_c(3900)}

\newcommand{\be}{\begin{equation}}
\newcommand{\ee}{\end{equation}}

\newcommand{\bqa}{\begin{eqnarray}}
\newcommand{\eqa}{\end{eqnarray}}

\begin{document}
\title{The $\zc$ peak does not come from the ``triangle singularity"}

\author{ Qin-Rong~Gong$^{1}$, Jing-Long~Pang$^{1}$, Yu-Fei~Wang$^{1}$, Han-Qing~Zheng$^{1,2}$
\vspace*{0.3cm} \\
$^{1}${\it Department of Physics and State Key Laboratory of Nuclear Physics and Technology,} \\
{\it Peking University,  Beijing 100871, China}\\
$^{2}${\it  Collaborative Innovation Center of Quantum Matter, Beijing 100871, China}
}

\maketitle

\begin{abstract}
 We  compare  contributions from the triangle diagram and  the $D\bar D^*$ bubble chain to the processes of $e^{+}e^{-}\rightarrow J/\psi\pi^{+}\pi^{-}$ and  $e^{+}e^{-}\rightarrow (D\bar D^*)^\mp\pi^{\pm}$. By fitting the $J/\psi\pi$ maximum spectrum and the $D\bar D^*$ spectrum, we find that the triangle diagram  cannot explain the new experimental results from BESIII Collaboration at center of mass at 4.23GeV and 4.26GeV, simultaneously. On the contrary, the molecular assignment of $\zc$ gives a much better description.
\end{abstract}

\vspace{1cm}
\section{Introduction}
The charged charmonium-like state $\zc$ was observed in $J/\psi\pi^\pm$ mass spectrum by BES III Collaboration in $e^+e^-\rightarrow J/\psi\pi\pi$ process~\cite{Ablikim:2013mio}, and confirmed by Belle~\cite{Liu:2013dau} and CLEO~\cite{Xiao:2013iha} Collaborations in the same processes. Afterwards, it was also observed in the $(D\bar D^*)^\pm$ invariant mass spectrum in the process of $e^{+}e^{-}\rightarrow D\bar D^*\pi^\mp$, and the quantum number of $\zc$ was determined to be $\mbox{I}(J^P)=1(1^+)$ by angular distribution analysis on $\pi\zc$ system~\cite{Ablikim:2013xfr}.
The experimental discovery has stimulated a lot of discussions because of the unique nature of $\zc$, as it could be the first unambiguous candidate of the long wanted tetra-quark state.

In a recent paper~\cite{Gong:2016hlt}, we have made a detailed  comparison between the $D\bar D^*$ molecule picture and the ``elementary'' picture, and concluded that $\zc$ is of $D\bar D^*$ molecular nature, using the pole counting method~\cite{Morgan:1992ge}.

However, it is also found in the literature another possible mechanism, called anomalous triangle singularity (ATS), to explain the singularity structure at $\zc$. ATS refers to a branch cut in a three point loop function other than the normal threshold. The study of ATS can be traced back to about 60 years ago. In Ref.~\cite{mandelstam} S. Mandelstam worked out the ATS branch point and discussed its effects on the deuteron electromagnetism form factor, and in Ref.~\cite{landau} L. D. Landau applied Landau equations to triangle diagrams to analyse ATS. Extensive studies on the triangle singularity using dispersion techniques can also be found in Ref.~\cite{newref}. Especially in the paper by Lucha, Melikhov and Simula of Ref.~\cite{newref}, a detailed dispersive analysis is given on different variables.

ATS has attracted renewed interests recently because it may contribute to peaks in some certain invariant-mass spectrums. In other words, some so-called ``exotic hadron states'' could be just the ATS peak rather than real particles; or even if real exotic hadron states exist, there may be some non-negligible contributions from ATS. For example, it is suggested in Refs.~\cite{wangqian,wuxg,Liu:2013vfa}
that the singularity structure of the triangle diagram (see Fig.~\ref{sandianhanshu}), which contains both the normal threshold effect and anomalous threshold effect, may lead to the peak at $3900$ MeV. In Refs.~\cite{Liu:2015taa,adam} it is emphasized that the anomalous triangle singularity may have significant impact in understanding the nature of the near-threshold state.

This paper devotes to the study of triangle diagram contribution to the $\zc$ peak. In Sec.~\ref{theory} we give a pedagogical analysis on general three point loop functions using Feynman parameter representation that can be found in most textbooks, and discuss the properties of the ATS. In Sec.~\ref{phenomenon} we calculate the triangle diagram corresponding to $e^{+}e^{-}\rightarrow J/\psi\pi^{+}\pi^{-}$ and $e^{+}e^{-}\rightarrow (D\bar D^*)^\mp\pi^{\pm}$ processes and fit the experimental data to test whether the $\zc$ peak comes from triangle diagram. In Sec.~\ref{conclusion} some conclusions are drawn. Basically, it is found that the new experimental results from Ref.~\cite{besnew} play a crucial role in clarifying the issue on triangle diagram contribution: the experimental data indicate that the peak at $4.23$ GeV is higher than that at $4.26$ GeV, whereas the triangle diagram predicts an opposite behavior. Our analysis reveals that $\zc$ peak cannot be explained from the triangle diagram contribution from Fig.~\ref{sandianhanshu}. Hence, combining with our previous analysis in Ref.~\cite{Gong:2016hlt}, the molecular nature of $\zc$
is firmly established.
\section{Theoretical framework}\label{theory}
\begin{figure}[htbp]
\begin{center}
\includegraphics[width=0.37\textwidth]{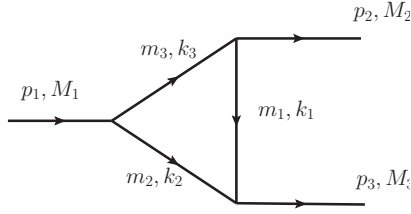}
\caption{\label{sandianhanshu} A general triangle diagram. The external and internal momenta are denoted as $p_i$ and $k_i$, and the internal and external masses are labeled as $m_i$ and $M_i$($i=1,2,3$), respectively.}
\end{center}
\end{figure}

To start let us first look upon a general triangle diagram as shown in Fig.~\ref{sandianhanshu}.
{ The amplitude for such a triangle diagram contains the following scalar three-point function:
\begin{align}
T(s_1,s_2,s_3)=-i\int\frac{d^4q}{(2\pi)^4}\frac{1}{(k_1^2-m_1^2+i \epsilon)(k_2^2-m_2^2+i \epsilon)(k_3^2-m_3^2+i \epsilon)}\ ,
\end{align}
where $k_1=q-p_2,k_2=p_1-q,k_3=q$, $s_i=p_i^2(i=1,2,3)$, and $\epsilon\to 0^+$ is a small positive parameter. This scalar three-point function could be evaluated using the standard method as follows. First one can rewrite it in terms of Feynman parameters $x,y$ and $z$
\begin{align}
T(s_1,s_2,s_3)=-i\int_0^1dxdydz\int\frac{d^4q}{(2\pi)^4}
\frac{2\delta(x+y+z-1)}{\big[x(k_1^2-m_1^2)+y(k_2^2-m_2^2)+z(k_3^2-m_3^2)+i(x+y+z)\epsilon\big]^3}\ .
\end{align}
After straightforward calculation, the function can be presented as
\begin{align}
T(s_1,s_2,s_3)=-i\int_0^1dx\int_0^{1-x}dy\int\frac{d^4q}{(2\pi)^4}
\frac{2}{\big[(q+xp_2-yp_1)^2-\Delta(x,y)+i \epsilon\big]^3}\ ,
\end{align}
where
\begin{align}
\Delta(x,y)=x^2s_2+y^2s_1+xy(s_1+s_2-s_3)-x(s_2-m_1^2+m_3^2)-y(s_1-m_2^2+m_3^2)+m_3^2\ .
\end{align}
One can perform a momentum translation $q=l-xp_2+yp_1$, then
\begin{align}
T(s_1,s_2,s_3)=-i\int_0^1dx\int_0^{1-x}dy\int\frac{d^4l}{(2\pi)^4}
\frac{2}{\big[l^2-\Delta(x,y)+i \epsilon\big]^3}\ .
\end{align}
To evaluate that integration, Wick rotation is done\footnote{Due to the $+i \epsilon$ term, poles in the $l^0$ plane, $l^0=\pm\sqrt{\Delta+|\vec{l}|^2-i \epsilon}$, always locate in the second and fourth quadrant, so Wick rotation is valid for both $\Delta\geq0$ and $\Delta<0$. }: $l^0\to i l^0_E, l^2\to -l^2_E$
\begin{align}
T(s_1,s_2,s_3)=-\int_0^1dx\int_0^{1-x}dy\int\frac{d^4l_E}{(2\pi)^4}
\frac{2}{\big[l_E^2+\Delta(x,y)-i \epsilon\big]^3}\ .
\end{align}
Then the integration in momentum space can be worked out anyway regardless of the sign of $\Delta(x,y)$\footnote{When $\Delta(x,y)<0$, the $-i \epsilon$ term guarantees the validity of this integration, see for example, M. E. Pesking and D. V. Schroeder, {\it An Introduction to Quantum Field Theory}, Westview Press, 1995, Page 808. }
\begin{align}
T(s_1,s_2,s_3)=-\frac{1}{16\pi^2}\int_0^1dx\int_0^{1-x}dy
\frac{1}{\Delta(x,y)-i \epsilon}\ .
\end{align}
Further, the integration of $y$ is then calculated analytically (for simplicity we omit the $-i\epsilon$ in the following discussion)
\begin{align}\label{Txint}
T(s_1,s_2,s_3)&=-\frac{1}{8\pi^2}\int_0^1 \frac{dx}{a(s_3,x)}
\Big\{\arctan\left[\frac{(-s_1+s_2-s_3)x+s_1+m_2^2-m_3^2}{a(s_3,x)}\right]\\
&-\arctan\left[\frac{(s_1+s_2-s_3)x-s_1+m_2^2-m_3^2}{a(s_3,x)}\right]\Big\}\nonumber\ ,
\end{align}
with
\begin{align}\label{Txa}
a(s_3,x)=&\Big\{ \big[\lambda(s_1,s_2,s_3)+\lambda(s_1,m_2^2,m_3^2)+4s_1 m_1^2-(s_2-s_3+m_2^2-m_3^2)^2\big]x\\
&-\lambda(s_1,s_2,s_3)x^2-\lambda(s_1,m_2^2,m_3^2)\Big\}^{1/2}\ ,
\end{align}
where
\begin{align}
\lambda(a,b,c)=a^2+b^2+c^2-2ab-2bc-2ac\ .
\end{align}

Now it's not difficult to analyse the singularity structures of $T(s_1,s_2,s_3)$ based on Eq.~\ref{Txint} and \ref{Txa}, and for simplicity we only study the singularities of $s_3$ variable. There are two ``$\arctan$'' terms in Eq.~\ref{Txint} ; the singularity of the first term is given by the $\pm i$ branch points of $\arctan$ function, i.e.
\[
\frac{(-s_1+s_2-s_3)x+s_1+m_2^2-m_3^2}{a(s_3,x)}=\pm i\ .
\]
Its solution with respect to the integration variable $x$ is
\begin{align}\label{xnormal}
x= \frac{-m_1^2+m_2^2+s_3\pm\sqrt{\lambda(s_3,m_1^2,m_2^2)}}{2s_3} \ .
\end{align}
If $\lambda(s_3,m_1^2,m_2^2)\to 0^-$, pinch singularity happens. Especially when $s_3\to (m_1+m_2)^2-0^+$, two singularities from Eq.~\ref{xnormal}
\[
x=\frac{m_2}{m_1+m_2}\pm i0^+
\]
will pinch the integration interval $[0,1]$ while $a(s_3,x)$ stays finite, hence it's reasonably concluded that $s_3=(m_1+m_2)^2$ is a singularity of $T(s_1,s_2,s_3)$\footnote{In fact there is another solution corresponding to $\lambda(s_3,m_1^2,m_2^2)\to 0^-$, that is, $s_3=(m_1-m_2)^2$, which is called pseudo-threshold and only appears on the un-physical sheet. So it's less relevant to our discussion. }. In fact, that result is nothing but the well known normal threshold of the three point function. One can do analogous analyses on the second $\arctan$ term of Eq.~\ref{Txint}, and it's found that the solution of $x$
\[
x=\frac{-m_1^2+m_3^2+s_2\pm\sqrt{\lambda(m_1^2,s_2,m_3^2)}}{2s_2}
\]
is independent of $s_3$, so the second term has no contribution to the singularities of $T(s_1,s_2,s_3)$ with respect to variable $s_3$.

On the other hand, if the singularities occur in the denominator $a(s_3,x)=0$, then the solution for $x$ variable is
\begin{equation}\label{pinchpoints}
\begin{split}
&x=\frac{-N(s_1,s_2,s_3)\pm\sqrt{N^2(s_1,s_2,s_3)-\lambda(s_1,m_2^2,m_3^2)\lambda(s_1,s_2,s_3)}}{\lambda(s_1,s_2,s_3)}\ ,\\
&N(s_1,s_2,s_3)=\lambda(s_1,s_2,s_3)+\lambda(s_1,m_2^2,m_3^2)+4s_1 m_1^2-(s_2-s_3+m_2^2-m_3^2)^2\ .
\end{split}
\end{equation}
When $s$ approaches the following two points,
\begin{align}\label{s3pm}
s^{\pm}=&\frac{1}{2 m_3^2}\bigg[2m_3^2(s_1+s_2)-(s_1-m_2^2+m_3^2)(s_2-m_1^2+m_3^2)\nonumber\\
&\pm\sqrt{\lambda(s_2,m_1^2,m_3^2)\lambda(s_1,m_2^2,m_3^2)}\bigg]\ ,
\end{align}
a pinch singularity occurs in $T(s_1,s_2,s_3)$. However, that singularity may not always appear on the physical Riemann sheet (i.e. the first sheet). Actually there exist two cases leading the singularity to be on the un-physical (the second) sheet: firstly, when the pinch points in Eq.~\ref{pinchpoints} lie off the integral interval $[0,1]$; secondly, when the numerator of the integrand in Eq.~\ref{Txint} approaches $0$ simultaneously as $a(s_3,x)\to 0$, giving a well defined value of the integrand. We will meet examples of these two cases later.

To proceed, we in the following discuss two situations, one is that all particles are stable, the other is that some particles are unstable like the case we meet in the real situation ($X(4260)\to J/\Psi \pi \pi, D D^* \pi$).

For the first situation, $(m_i-m_k)^2<s_j<(m_i+m_k)^2$ ($i,j,k$ are the permutation of 1,2,3). When we substitute the $s^+$ in Eq.~\ref{s3pm} for $s_3$ in Eq.~\ref{pinchpoints}, it will be found that $x$ is not in the integral interval, thus  $s^+$ locates on the second sheet. But the situation is more complicated when we focus on $s^-$, which is usually called ``anomalous threshold'' or ``anomalous triangle singularity''. Since the pinch points in Eq.~\ref{pinchpoints} stay in $[0,1]$ when $s\to s^-$, one has to test whether the numerator of the integrand in Eq.~\ref{Txint} approaches $0$. We have the following observations:
\begin{itemize}
\item when $s_1<(m_2+m_3)^2+\frac{m_2}{m_1}[(m_3-m_1)^2-s_2]$, the $s^-$ locates on the second sheet and is below the normal threshold;

\item when $s_1=(m_2+m_3)^2+\frac{m_2}{m_1}[(m_3-m_1)^2-s_2]$, the $s^-$ just rides on the normal threshold;

\item when $s_1>(m_2+m_3)^2+\frac{m_2}{m_1}[(m_3-m_1)^2-s_2]$, the $s^-$ locates on the first sheet and is below the normal threshold.
\end{itemize}
To obtain above results, one needs to analyse the behaviour of the numerator of the integral in Eq.~\ref{Txint} at $s^-$ by looking upon the arguments of the two $\arctan$ functions. For example, when $s_1<(m_2+m_3)^2+\frac{m_2}{m_1}[(m_3-m_1)^2-s_2]$, it is found that} {the first argument
\[
\lim_{s_3\to s^-}\lim_{x\to x_1}\frac{(-s_1+s_2-s_3)x+s_1+m_2^2-m_3^2}{a(s_3,x)}=-\infty\ ,
\]
while the second
\[
\lim_{s_3\to s^-}\lim_{x\to x_1}\frac{(s_1+s_2-s_3)x-s_1+m_2^2-m_3^2}{a(s_3,x)}=-\infty\ ,
\]
using $\arctan(-\infty)=-\pi/2$, one finds the two $\arctan$ terms cancel each other in Eq.~\ref{Txint} when $s_3\to s^-$, so in this case $s^-$ is not on the physical sheet.} { But when $s_1>(m_2+m_3)^2+\frac{m_2}{m_1}[(m_3-m_1)^2-s_2]$ the first argument becomes $+\infty$, leading the numerator to be nonzero
\[
\lim_{s_3\to s^-}\lim_{x\to x_1} \arctan[\cdots]-\arctan[\cdots]=\pi/2-(-\pi/2)=\pi\ ,
\]
thus $s^-$ appears on the first sheet. In above discussions the behavior of $s^-$ is actually the classical example discovered long time ago by Mandelstam in Ref.~\cite{mandelstam}, being used to explain the long tail of deuteron wave function.

In reality, however, the stability condition in obtaining the above results may not hold. For example, one may consider the following kinematics (which corresponds to the kinematics of Fig.~\ref{triangle}):
\begin{align}\label{kinematics}
0<s_2<(m_1-m_3)^2\ ,\,\,\,(m_1+m_2)^2<s_3<(\sqrt{s_1}-\sqrt{s_2})^2\ .
\end{align}
After some analyses similar to the situation of stable particles, we find that when $s_1$ satisfies
\begin{align}\label{kinematics2}
(m_2+m_3)^2 <s_1<(m_2+m_3)^2+\frac{m_2}{m_1}[(m_3-m_1)^2-s_2]\ ,
\end{align}
$s^-$ locates on the physical sheet, and is above the normal threshold. {Otherwise}, $s^{-}$ would be on the un-physical sheet.

In addition, to understand the dependence of $s^-$ on $s_1$,
one gives $s_1$ a small positive imaginary part, $s_1\rightarrow s_1+i0^+$. Then $s^-$ can be expressed as $s^-(s_1+i0^+)=s^-(s_1)+i\frac{\partial s^-}{\partial s_1}0^+$ derived from Eq.~\ref{s3pm}. As $s_1$ increases, the near-threshold trajectory of $s^-$ both in classical stable case (see Fig.~\ref{fig:subfig:a1}) and under the kinematics of Eq.~\ref{kinematics} (see Fig.~\ref{fig:subfig:b1}) can thus be drawn.
\begin{figure}[htbp]
\center
\subfigure[]{
\label{fig:subfig:a1}
\scalebox{1.2}[1.2]{\includegraphics[width=0.35\textwidth]{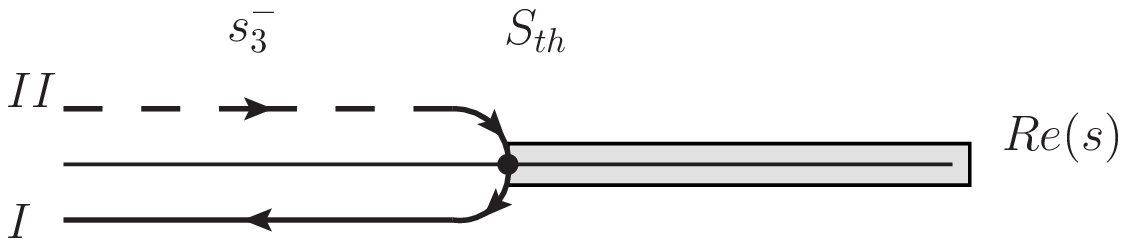}}}
\subfigure[]{
\label{fig:subfig:b1}
\scalebox{1.2}[1.2]{\includegraphics[width=0.35\textwidth]{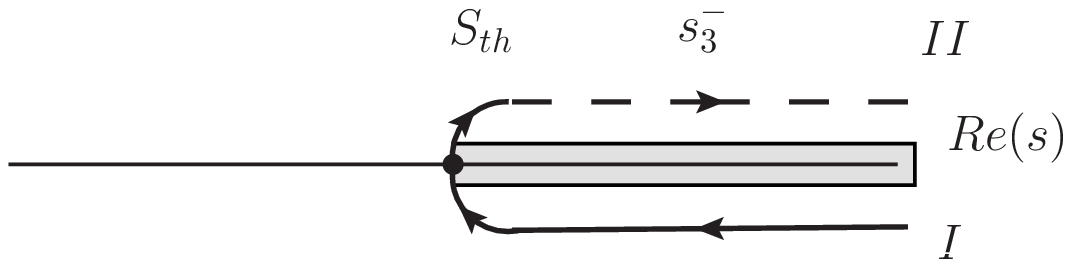}}}
\caption{The trajectory of the anomalous threshold $s^-$:
 (a) All particles are stable; (b) {The kinematics from Eq.~\ref{kinematics}.} }\label{tra}
\end{figure}

The aim of this paper is to investigate whether $Z_c(3900)$ peak is mainly from triangle singularity or not, so the processes $X(4260)\rightarrow\pi^{+}\pi^{-} J/\psi, D \bar{D}^* \pi$ are considered, with respect to the triangle diagrams shown in Fig.~\ref{triangle} {as suggested by Refs.~\cite{wangqian} -- \cite{adam}}. We set $s_1$ to be the square of $X(4260)$ $4-$ momentum, and the pertinent masses to the masses of those particles, then according to Eq.~\ref{kinematics2}, it's found that the ATS lies on the second sheet when $\sqrt{s_1}$ lies between $4230$MeV and $4260$MeV, hence the $\zc$ peak cannot be a direct manifestation of the anomalous threshold. We plot the modulus-square of the amplitude in Eq.~\ref{Txint} with different center of mass energies $\sqrt{s_1}$, as shown in Fig.~\ref{amp1}.

It is however found that the location and effect of the anomalous threshold are very sensitive to the energy of $X(4260)$.
When the anomalous threshold is on the second sheet as shown in {Fig.} \ref{fig:subfig:a11}, the closer it is to normal threshold the more influence it has on the amplitude.  Since the anomalous threshold (on the second sheet) can be rather close to the normal threshold, one still needs to check whether the (anomalous and normal) threshold effects can cause the  experimentally observed $Z_c(3900)$ peak.
\begin{figure}[htbp]
\center
\subfigure[]{
\label{fig:subfig:a11}
\scalebox{1.2}[1.2]{\includegraphics[width=0.37\textwidth]{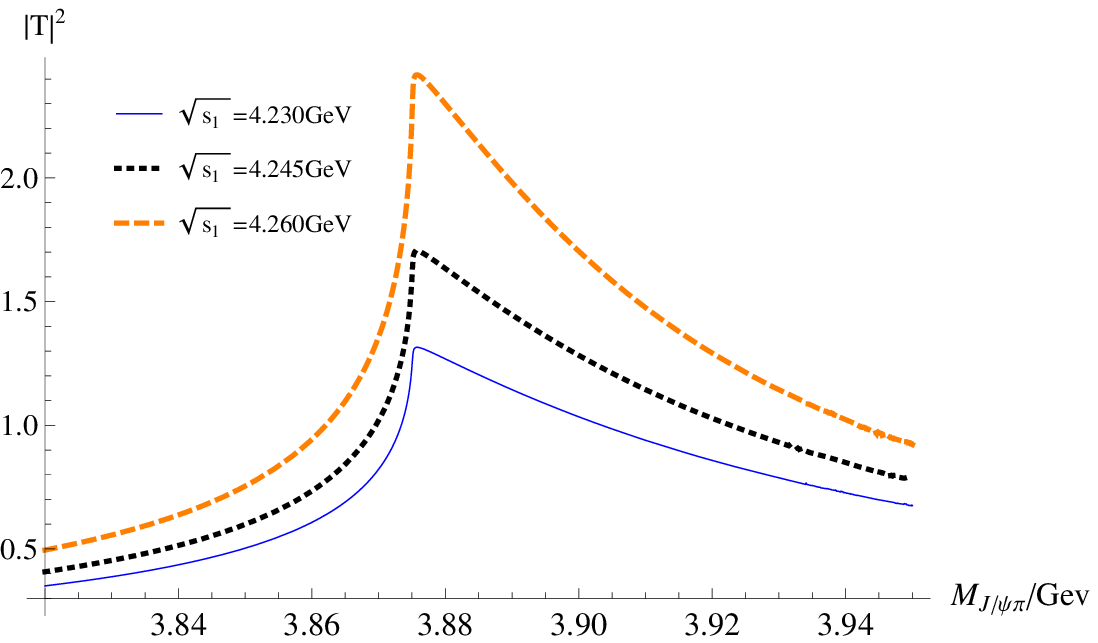}}}
\subfigure[]{
\label{fig:subfig:bb}
\scalebox{1.2}[1.2]{\includegraphics[width=0.37\textwidth]{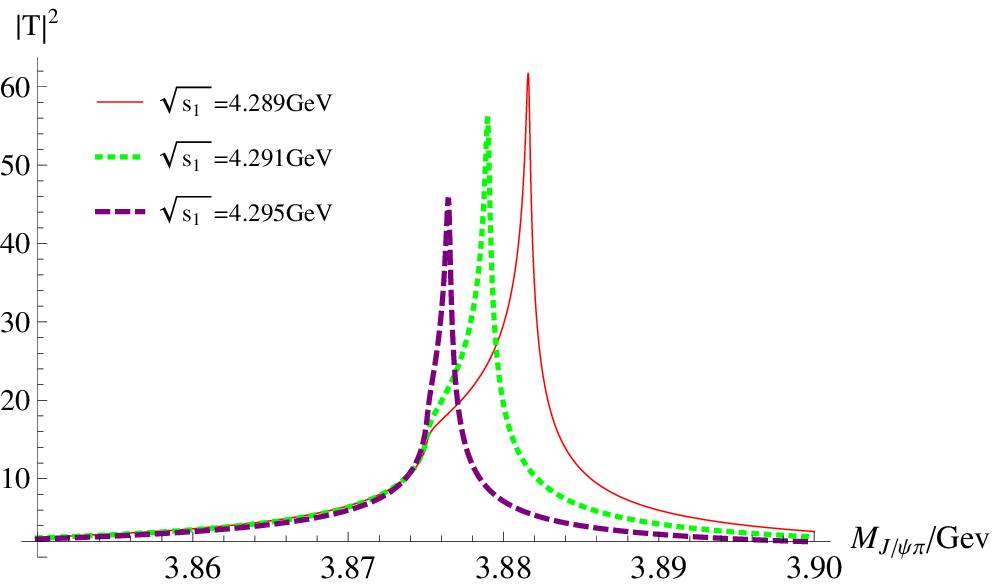}}}
\caption{Invariant mass distributions of the process $X(4260)\rightarrow\pi\pi J/\psi$ from the triangle diagram for different $\sqrt{s_1}$:
 (a) The ATS is on the second sheet; (b) The ATS is on the first sheet, and above the normal threshold.}\label{amp1}
\end{figure}

{The discussions above are only brief qualitative analyses aiming at studying the dependence of ATS peak on $s_1$ variable, and what we have calculated above is not the whole amplitude of that process to fit experimental data, since in the full amplitude there exists $X(4260)$ production process, $X(4260)$ propagator, spin structures, as well as derivative couplings. The detailed formulae can be found in Appendices.~\ref{appe.A} and \ref{appe.B}, with which the numerical discussions are made in the next section. }
\section{Numerical Analyses and Discussions}\label{phenomenon}
\subsection{Fit to the data of Ref.~\cite{Ablikim:2013mio,Ablikim:2013xfr}}
As pointed out in Ref.~\cite{Liu:2015taa},
the ATS contribution to the decay of $X(4260)$ to $J/\psi\pi\pi$ (see Fig.~\ref{fig:subfig:bb}) may have great dependence on the center-of-mass energy. However Ref.~\cite{Liu:2015taa} just made a rough numerical discussion.

On the other side, the result in Ref.~\cite{Gong:2016hlt} supports molecular state interpretation rather than ``elementary'' state explanation. Since the aim of this paper is to test whether the triangle diagram can provide the $\zc$ peak,  we make three independent fits to compare the fit quality: 
\begin{itemize}
  \item{Fit I: $X(4260)$ decays to final states only through triangle diagram as depicted in Fig.~\ref{triangle}.}
  \item{Fit II: the final states are only produced by $D\bar {D}^*$ rescattering, 
   and the Feynman diagram is shown in Fig.~\ref{bubble}.}
  \item{{Fit III: the mixed situation by combining the triangle diagram and $D\bar D^*$ bubble chains, as shown in Fig.~\ref{mixed}. }}
\end{itemize}
Recall that both the bubble and the triangle diagrams are ultraviolet divergent. We use $\overline{\text{MS}}-1$ scheme of dimensional renormalization method to deal with the divergences, which leaves another somewhat arbitrary parameter: the renormalization scale $\mu$ {(see Appendix.~\ref{appe.B})}. Considering the physical process we are studying, it is reasonable to expect that the $\mu$ parameter should be around or not much differ from the mass of $X(4260)$. Indeed, we find that when we set the renormalization scale at a reasonable value $\mu=5$GeV, the bubble chains give a satisfactory fit result as shown in the following text, and it's verified the fit quality is not sensitive to the variation of $\mu$ parameter. On the contrary, the triangle diagram does not give a good description to the data when $\mu$ is set at $5$GeV. The fit results shown in the following text correspond to an unreasonably small $\mu$ (in getting the Fig.~\ref{fit2} and Fig.~\ref{newfit}). \footnote{This fact actually indicates that the triangle diagram does not work in simulating the $Z_c$ peak. }

In the numerical fit two sets of data including $J/\psi\pi$ maximum invariant mass spectrum~\cite{Ablikim:2013mio} and $D\bar D^*$ invariant mass distribution~\cite{Ablikim:2013xfr} are taken into account.

\begin{figure}[htbp]
\begin{center}
\includegraphics[width=0.65\textwidth]{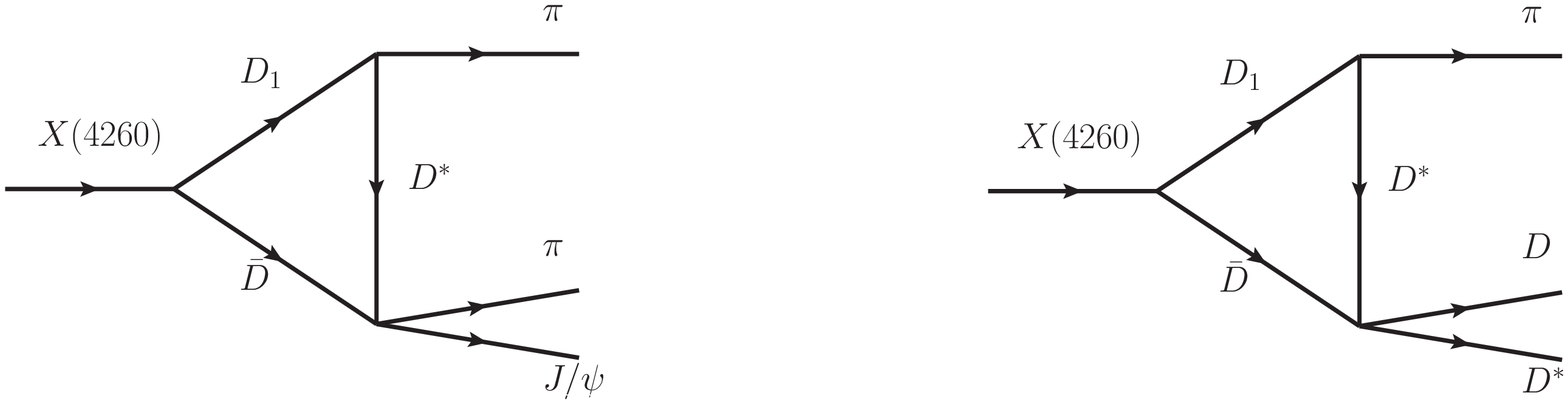}
\caption{Feynman diagrams of Fit I.}\label{triangle}
\end{center}

\end{figure}
\begin{figure}[htbp]
\begin{center}
\includegraphics[width=0.8\textwidth]{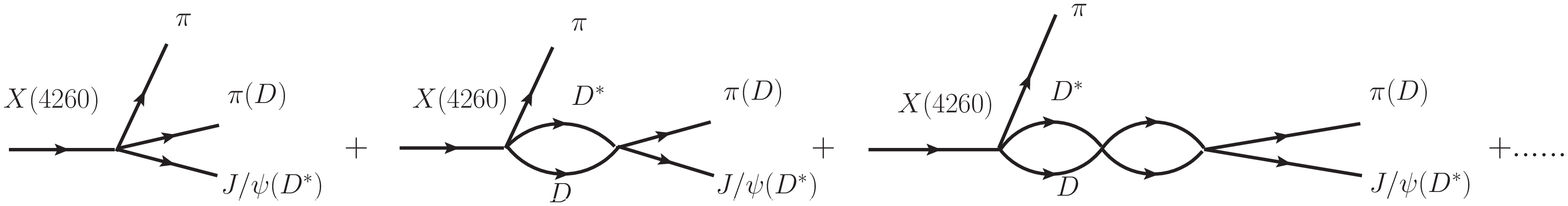}
\caption{Feynman diagrams of Fit II.}\label{bubble}
\end{center}

\end{figure}
\begin{figure}[htbp]
\begin{center}
\includegraphics[width=0.7\textwidth]{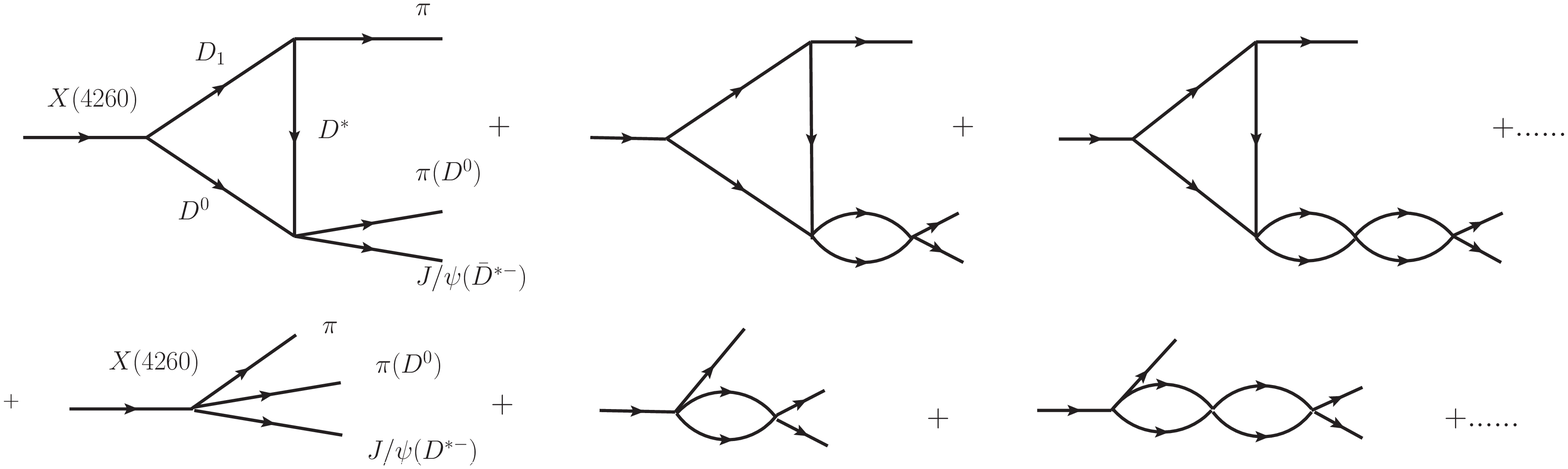}
\caption{ Feynman diagrams of Fit III. }\label{mixed}
\end{center}
\end{figure}

The Lagrangian we use is taken from Ref.~\cite{Gong:2016hlt} with three additional pieces£º
\be
\mathcal{L}_{XD_1D}=h_1X_\mu \langle D_{1}^\mu\bar{D}+ h.c.\rangle\ ,
\ee
\be
\mathcal{L}_{D_1D^*\pi}=h_2\langle\bigtriangledown^\mu D_1^{\nu}\cdot\bar{D}^{*}_\nu u_\mu\rangle
\ee
and
\be
\mathcal{L}_{\gamma^*X}=c_\gamma F_{\mu\nu}X^{\mu\nu}\ ,
\ee
{where $X^{\mu\nu}=\partial^\mu X^\nu-\partial^\nu X^\mu$. The corresponding Feynman rules can be found in Appendix.~\ref{appe.A}. }

With these  preparations, it is possible to make a combined fit on the $J/\psi\pi$ maximum spectrum and $D\bar D^*$ mass distribution~\cite{Ablikim:2013mio,Ablikim:2013xfr}. Except for various coupling parameters, two parameters for the $D\bar D^*$ incoherent background and two normalization constants are further introduced. In total, there are 8 and 10 free parameters for Fit I and Fit II, respectively.

Since the value of the center-of-mass energy of $X(4260)/\gamma^*$  severely influences the ATS contribution as discussed previously, we in the fit also carefully analyze the effect of energy resolution both in  $X(4260)$ channel and in $\zc$ channel. However it is found that
the effect of the energy resolution does not obviously improve the fit quality, since the energy resolution parameters are much smaller than the particle widths.\footnote{G. Y.~Tang, private communications.} The fit results are shown in Fig.~\ref{fit2}. The $\chi^2/d.o.f$ of Fit I and Fit II are  2.5 and 0.96, respectively. In Fit II we find a bound state pole of $DD^*$, $\sqrt{s}=3.8747\pm 0.0148i$GeV, which is consistent with our previous result in Ref.~\cite{Gong:2016hlt}.
{Further,  Fit III gives very similar $\chi^2$ to Fit II with pole located at $\sqrt{s}=3.8749\pm0.0145i$GeV, which suggests that the triangle diagram plays only a minor role as comparing with the bubble chain contribution. } {In Fit II and Fit III, the denominators of amplitudes of the processes $e^{+}e^{-}\to J/\psi\pi\pi(D\bar D^*\pi)$ take the form
\begin{align}
  1-i\lambda_1 (D(l)+c_0),
\end{align}\label{eqpole}
where the $\lambda_1$ represents the $D\bar D^*\bar DD^*$ contact coupling constant, and $c_0$ simulates the contribution from other lighter channels to the $\zc$ width. The function $D(l)$ is the $D\bar D^*$ meson loop integral.
The parameters $\lambda_1$ and $c_0$ which decide the pole positions are listed in Table~\ref{table1}.\footnote{Other coupling constants and the normalization constants are multiplied to each other and are not quite interesting physically, so we do not list them here.}}
\begin{table}[htbp]
\begin{center}
 \begin{tabular}  {||c| c | c|c ||}
 \hline
    & $\lambda_1$ & $c_0$ & pole position (GeV)\\
   \hline
 Fit II &$-345.29$ & $0.002347$ & $3.8747\pm0.0148i$\\
  \hline
  Fit III&$-341.23$ & $0.002342$ & $3.8749\pm0.0145i$\\
  \hline
 \end{tabular}\\
 \caption{ The parameters which decide the pole positions of Fit II and Fit III.
 }\label{table1}
   \end{center}
\end{table}
\begin{figure}[htbp]
\center
\subfigure[]{
\label{fig:subfig:olda}
\scalebox{1.2}[1.2]{\includegraphics[height=1.5in,width=2.3in]{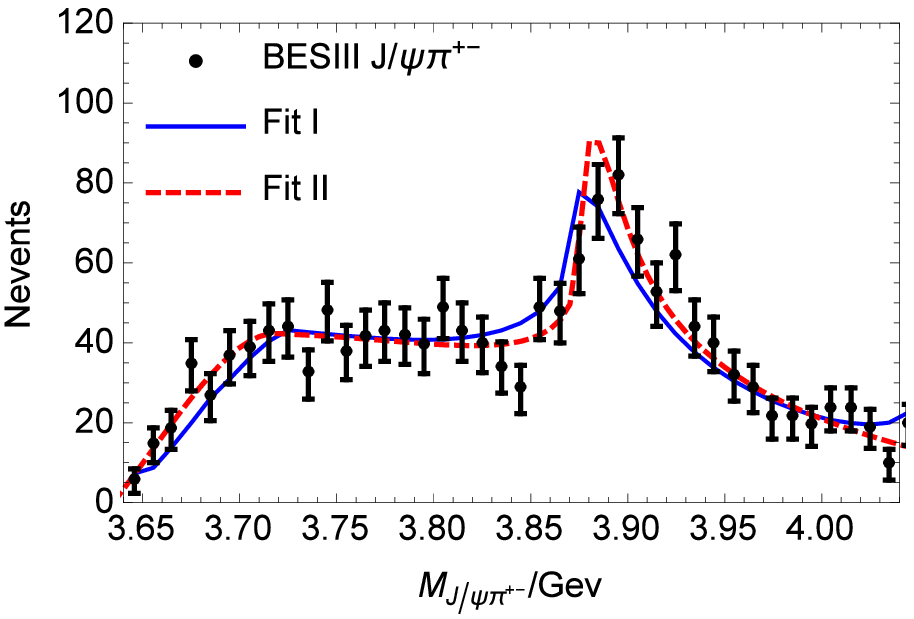}}}
\subfigure[]{
\label{fig:subfig:oldb}
\scalebox{1.2}[1.2]{\includegraphics[height=1.5in,width=2.3in]{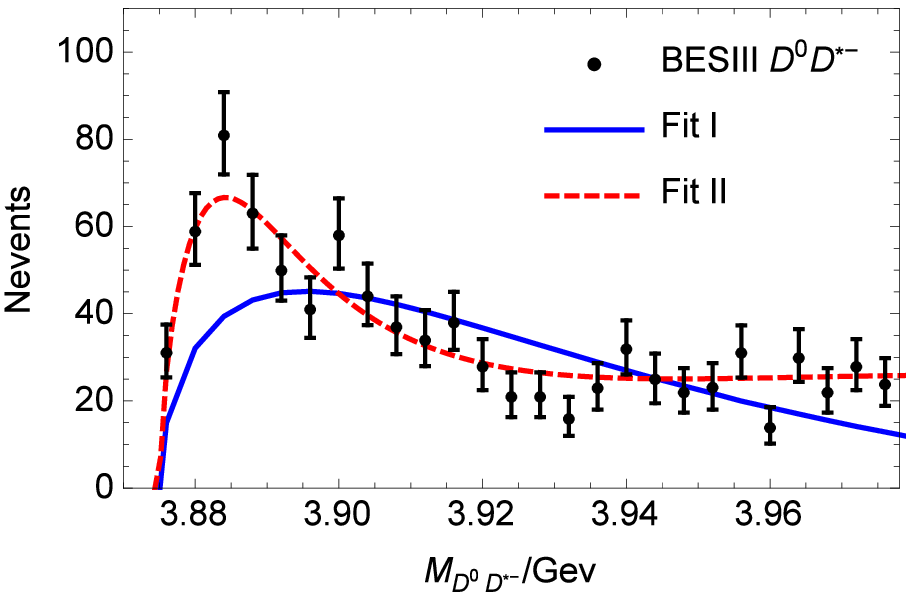}}}
\caption{Data fit using two different parameterizations (Fit I, II) of the amplitudes. (a): the $J/\psi\pi$ maximum invariant mass distribution from Ref.~\cite{Ablikim:2013mio}; (b): the $D\bar D^*$ invariant mass distribution from Ref.~\cite{Ablikim:2013xfr}.}\label{fit2}
\end{figure}

Although the bubble chains (the molecule picture) fit data  better than the triangle diagram, the latter still cannot be firmly excluded here. Generally speaking, one believes that the cusp effects are much weaker singularities than poles, and the triangle diagram does not generate poles except  branch point singularities. However, the undetermined overall normalization factor in the present fit makes up the defect of branch cuts (i.e., being weak in general), and hence prevents us from excluding the triangle diagram mechanism.
\subsection{Fit to the new data}
Fortunately, the new data from BESIII Collaboration~\cite{besnew}  indicates that there are more events in $\zc$ peak at $\sqrt{s_1}=4.23$GeV (with a integrated luminosity of $L=1092$ pb$^{-1}$) than at $\sqrt{s_1}=4.26$GeV (with $L= 827$ pb$^{-1}$), after background subtraction (see Fig.~\ref{newfit}).
On the contrary, the magnitude of the triangle diagram in Fig.~\ref{triangle}  at the $Z_c$ peak is smaller when $\sqrt{s_1}=4.23$GeV comparing with the magnitude when $\sqrt{s_1}=4.26$GeV.\footnote{Since the latter is closer to the $\bar DD_1$ threshold.}   It is noticed that the $s_1$ dependence of the triangle diagram will be slightly balanced
by the $s_1$ dependence of $X(4260)$ propagator, which takes the standard Breit-Wigner form of constant width taken from PDG.\footnote{Other choice of $X(4260$ propagator like the one in Ref.~\cite{tgy2015} leads to very similar results. In Refs.~\cite{wangqian,wuxg,Liu:2013vfa} it is suggested that $X(4260)$ is a $D_1\bar D$ molecule, which will however make Fit I' even worse.} On the other side, different from the triangle diagram, the bubble chain amplitude is not sensitive to $s_1$ in the energy region of interests.
New fits to both the $\sqrt{s_1}=4.23$GeV and the $\sqrt{s_1}=4.26$GeV data are performed.

The results are shown in Fig.~\ref{newfit}. The $\chi^2_{dof}=5.3$ for pure triangle diagram (Fit I') and the $\chi^2_{dof}=1.6$ for pure bubble resummation (Fit II'). {The pole of Fit II' locates at $\sqrt{s}=3.8804\pm0.0150i$GeV.} Hence the triangle diagram gives a much worse fit comparing with the bubble chain diagram, and hence can be ruled out. Apparently, the new data \cite{besnew} are crucial in supporting of the $D\bar D^*$ molecule explanation of $\zc$.

\begin{figure}[htbp]
\center
\subfigure[]{
\label{fig:subfig:a}
\scalebox{1.2}[1.2]{\includegraphics[height=1.5in,width=2.3in]{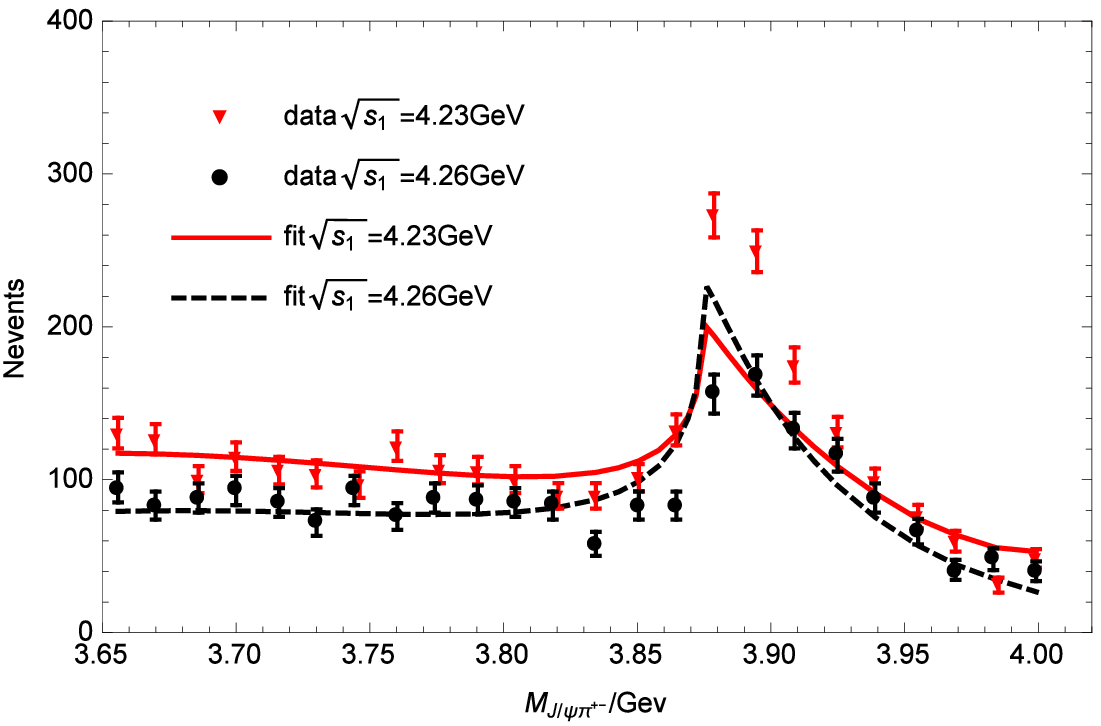}}}
\subfigure[]{
\label{fig:subfig:b}
\scalebox{1.2}[1.2]{\includegraphics[height=1.5in,width=2.3in]{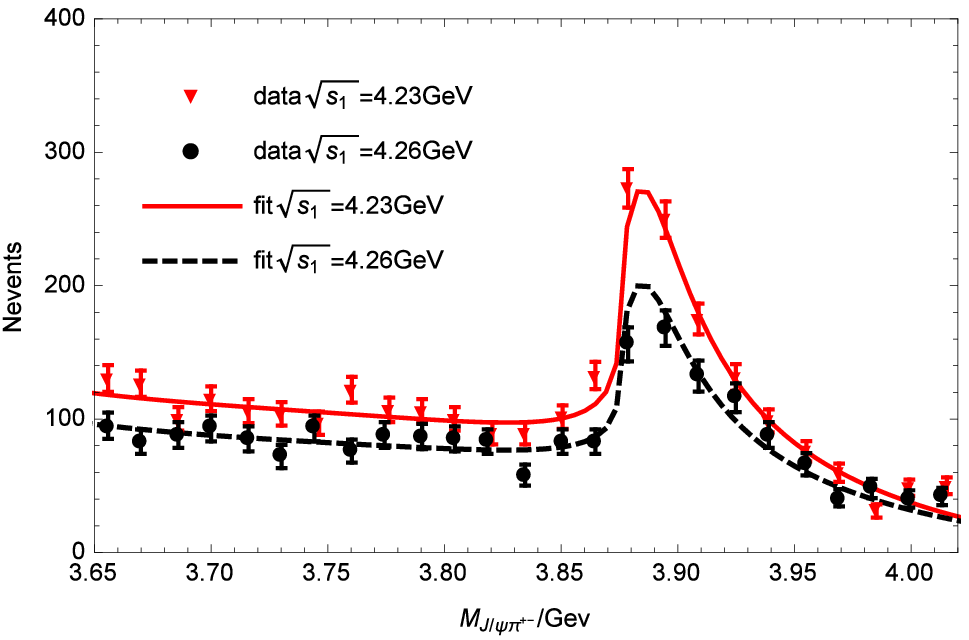}}}
\caption{Simultaneous fit to data at $\sqrt{s_1}=4.23$GeV and data at $\sqrt{s_1}=4.26$GeV. The integrated luminosity of these two data sets are renormalized to be equal. (a): Fit I', triangle diagram;\,\,\, (b): Fit II', bubble chains. Data are from Ref.~\cite{besnew}.}\label{newfit}
\end{figure}

{We also fit the new data of Fig.~\ref{newfit} in the mixed situation (Fit III'), however it does not obviously improve the total $\chi^2$ comparing with Fit II'. I.e., the $\chi^2_{d.o.f.}$ is almost the same as Fit II', with the pole location
$\sqrt{s}=3.8822\pm0.0119i$GeV, see Table~\ref{table2}.
Hence we may draw the conclusion that the bubble chain mechanism plays a dominant role in reproducing the experimentally observed peak structure. }

\begin{table}[htbp]
\begin{center}
 \begin{tabular}  {||c| c | c |c ||}
 \hline
    & $\lambda_1$ & $c_0$  & pole position (GeV)\\
   \hline
 Fit II' &$-305.48$ & $0.002871$ & $3.8804\pm0.0150i$\\
  \hline
  Fit III'&$-271.10$ & $0.002859$ & $3.8822\pm0.0119i$\\
  \hline
 \end{tabular}\\
 \caption{ The parameters which decide the pole position of Fit II' and Fit III'. }\label{table2}
   \end{center}
\end{table}

Here we should mention that in the above fit using triangle diagrams, the renormalization scale $\mu$ runs to a ridiculously small number, $\sim 10^{-7}$ GeV. If we fix $\mu$ at $5$ GeV, the fit quality of triangle diagrams gets even worse.
\section{Conclusion}\label{conclusion}
To summarize, we have investigated whether the triangle singularity mechanism proposed in the literature can be responsible for the experimentally observed $\zc$ peak.
It is found that, though the triangle diagram could barely explain the line shape, up to an arbitrary normalization constant, it fails to explain the dependence of the process  $e^{+}e^{-}\rightarrow J/\psi\pi^{+}\pi^{-}$ on the center of mass energy, not to mention the weird value of the renormalization scale it requires. Therefore, we conclude that $\zc$ peak is dominantly contributed by the pole of the $D \bar D^*$ molecular state.

The authors are grateful to Chang-zheng Yuan for bringing Ref.~\cite{besnew} to our attention, and Guang-Yi Tang for valuable discussions. We also would like to thank Qiang Zhao and Feng-Kun Guo for helpful discussions.
This work is supported in part by  National Nature Science Foundations of China (NSFC) under Contract Nos. 10925522, 11021092.

{Note added: when this paper is being completed, we became aware of a recent paper~\cite{adam2}, where the authors also attacked the same problem using four different amplitude parameterizations. They reach a conclusion that at this stage they can not have a preference on one of these parameterizations, which contradicts our conclusion. We point out here that they were only able to use the old data (at c.m. energy $4.23$GeV) in the neutral channel of $J/\psi \pi^0$~\cite{besold}, which is much worse in statistics comparing with the data of~\cite{besnew}. Hence we urge the authors of Ref.~\cite{adam2} redo their analysis with the new data incorporated and compare with our result. }

\appendix{\center\bf\huge Appendix}
{
\section{The pertinent Feynman rules}\label{appe.A}
For charged final state $DD^*$ or $J/\psi\pi$, Feynman vertices in Fig.~\ref{triangle} are given as follows.
\begin{align}
&iV_{X^\mu DD_1^\nu}=i h_1 g^{\mu\nu}\ ,\\
&iV_{D_1^\mu D^{*\nu}\pi}=-\frac{i\sqrt{2}h_2}{f_\pi}g^{\mu\nu}(p_{D_1}\cdot p_{\pi})\ ,\\
&iV_{DD^{*\mu}J/\psi^\nu\pi}=
\frac{i\sqrt{2}}{f_\pi}\left[g^{\mu\nu}(\lambda_2p_{J/\psi}\cdot p_\pi-\lambda_3p_{D^*}\cdot p_\pi)+(\lambda_4p_{J/\psi}^{\mu} p_\pi^\nu-\lambda_5p_{D^*}^\nu p_\pi^{\mu})\right]\ ,\\
&iV_{DD^{*\mu}DD^{*\nu}}=2i\lambda_1g^{\mu\nu}\ ,
\end{align}
where $\mu, \nu$ are Lorentz indices of spin$-1$ particles, $f_\pi$ is the pion decay constant, the flat space-time metric is $g^{\mu\nu}=\text{diag}\{1,-1,-1,-1\}$, and $p_{A}$ denotes the four-momentum of particle $A$ in the vertex. As for propagators of $D^*,\ D_1$ and $J/\psi$, we adopt the standard formula of Proca fields with four-momentum $k$ mass $M_P$, i.e.
\begin{equation}
iD_{F\,\rho\sigma}(k)=\frac{-i(g_{\rho\sigma}-\frac{k_\rho k_\sigma}{M_P^2})}{k^2-M_P^2}\ .
\end{equation}
When one calculates the modulus-square of the amplitudes, the polarization summing of the final state $D^*$ and $J/\psi$ is needed, and the formula of Proca field physical polarization summing is as follows:
\begin{equation}
\sum\varepsilon^*_{\alpha}\varepsilon_{\beta}=-g_{\alpha\beta}+\frac{k_{\alpha}k_{\beta}}{M_P^2}\ ,
\end{equation}
with $\varepsilon(k)$ being the final state Proca field polarization vector and $k$ being the four-momentum of the Proca field.

The initial $e^+e^-\to \gamma^*(\mu)$ QED process is written as
\begin{equation}
\bar{v}(p_+,s)(-i e\gamma^\mu)u(p_-,r)\ ,
\end{equation}
with the initial positron $4-$ momentum being $p_+$ and polarization being $s$, and the electron's $p_-$ and $r$. Moreover, the $\gamma^*$ ($\mu$) and $X(4260)$ ($\nu$) two-point coupling is
\begin{align}
iV_{\gamma^{*\mu} X^\nu}=2i (g^{\mu\nu}p_1^2-p_1^\mu p_1^\nu)c_\gamma\ ,
\end{align}
with $p_1=p_++p_-$. That form ensures QED Ward identity.
\section{Amplitudes of the total process}\label{appe.B}
\begin{figure}[htbp]
\centering
\includegraphics[width=0.6\textwidth]{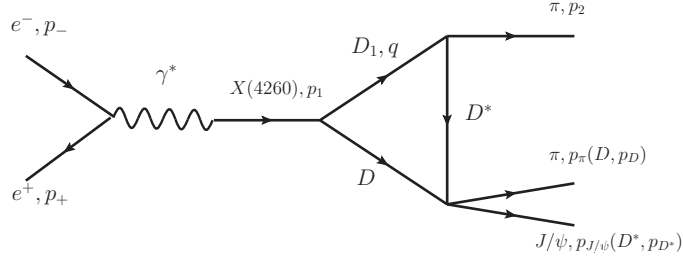}
\caption{The process of $e^+e^-\rightarrow J/\psi\pi\pi (D\bar{D}^*\pi)$.}
\label{e+e-}
\end{figure}
In this section we denote the $4-$ momentum of $X(4260)$ as $p_1$, the $4-$ momentum of final state $\pi$ in the $3-$ vertex as $p_2$, and $4-$ momentums of final state $\pi$ and $J/\psi$ in the $4-$ vertex as $p_\pi$ and $p_{J/\psi}$, respectively, and similar for the final state $D$ and $D^*$ (see Fig.~\ref{e+e-}). The total amplitudes with triangle diagrams of $DD^*\pi$ or $J/\psi\pi\pi$ final states can be written as
\begin{equation}
\begin{split}
i\mathcal{M}=\bar{v}(p_+,s)\gamma^\alpha u(p_-,r)\frac{2ie c_1(p_1^2g_{\alpha\mu}-p_{1\alpha}p_{1\mu})}{p_1^2 \big[p_1^2-M_X^2+iM_X\Gamma(p_1^2)\big]}(iT_{DD^*,J/\psi \pi}^{\mu\sigma})\cdot\varepsilon^*_\sigma(p_{D^*,J/\psi},f)
\ \mbox{, }
\end{split}
\end{equation}
while the triangle parts:
\begin{equation}\label{DDstartriampl}
\begin{split}
&iT^{\mu\sigma}_{DD^*}=\frac{-2\sqrt{2}\lambda_1 h_1h_2}{f_\pi}\\
&\times\int\frac{d^Dq\nu^{2\epsilon}}{(2\pi)^D}\frac{(q\cdot p_2)
\left\{g ^{\mu\sigma}-\frac{1}{M_{D_1}^2M_{D^*}^{2}}\Big[M_{D_1}^2(q-p_2)^\mu(q-p_2)^\sigma+M_{D^*}^{2}q^\mu q^\sigma-q\cdot(q-p_2)q^\mu(q-p_2)^\sigma\Big]\right\}}{(q^2-M_{D_1}^{2})\big[(q-p_1)^2-M_{D}^{2}\big]\big[(q-p_2)^2-M_{D^*}^{2}\big]}\ \mbox{, }
\end{split}
\end{equation}
\begin{equation}\label{Jpitriampl}
iT^{\mu\sigma}_{J/\psi\pi}=-\frac{2h_1h_2}{f_\pi^2}\left(\sum_{j=2}^{5}\lambda_jit^{\mu\sigma}_{\lambda_j}\right)\ \mbox{, }
\end{equation}
with
\begin{equation}
\begin{split}
it^{\mu\sigma}_{\lambda_2}&=\int\frac{d^Dq\nu^{2\epsilon}}{(2\pi)^D}(p_{J/\psi}\cdot p_{\pi})g^{\rho'\sigma}
\frac{g_{\rho\rho'}-\frac{(q-p_2)_\rho(q-p_2)_{\rho'}}{M_{D^*}^{2}}}{(q-p_2)^2-M_{D^*}^{2}}\left[g^{\rho\nu'}(q\cdot p_2)\right]\\
&\times \frac{g_{\nu'\nu}-\frac{q_\rho q_{\rho'}}{M_{D_1}^{2}}}{q^2-M_{D_1}^{2}}g^{\mu\nu}\frac{1}{(q-p_1)^2-M_{D}^{2}}\ \mbox{, }
\end{split}
\end{equation}
\begin{equation}
\begin{split}
it^{\mu\sigma}_{\lambda_3}&=-\int\frac{d^Dq\nu^{2\epsilon}}{(2\pi)^D}p_{\pi}\cdot(q-p_2)g^{\rho'\sigma}
\frac{g_{\rho\rho'}-\frac{(q-p_2)_\rho(q-p_2)_{\rho'}}{M_{D^*}^{2}}}{(q-p_2)^2-M_{D^*}^{2}}\left[g^{\rho\nu'}(q\cdot p_2)\right]\\
&\times \frac{g_{\nu'\nu}-\frac{q_\rho q_{\rho'}}{M_{D_1}^{2}}}{q^2-M_{D_1}^{2}}g^{\mu\nu}\frac{1}{(q-p_1)^2-M_{D}^{2}}\ \mbox{, }
\end{split}
\end{equation}
\begin{equation}
\begin{split}
it^{\mu\sigma}_{\lambda_4}&=\int\frac{d^Dq\nu^{2\epsilon}}{(2\pi)^D}p_{J/\psi}^{\rho'}p_{\pi}^\sigma
\frac{g_{\rho\rho'}-\frac{(q-p_2)_\rho(q-p_2)_{\rho'}}{M_{D^*}^{2}}}{(q-p_2)^2-M_{D^*}^{2}}\left[g^{\rho\nu'}(q\cdot p_2)\right]\\
&\times \frac{g_{\nu'\nu}-\frac{q_\rho q_{\rho'}}{M_{D_1}^{2}}}{q^2-M_{D_1}^{2}}g^{\mu\nu}\frac{1}{(q-p_1)^2-M_{D}^{2}}\ \mbox{, }
\end{split}
\end{equation}
\begin{equation}
\begin{split}
it^{\mu\sigma}_{\lambda_5}&=-\int\frac{d^Dq\nu^{2\epsilon}}{(2\pi)^D}p_{\pi}^{\rho'}(q-p_2)^\sigma
\frac{g_{\rho\rho'}-\frac{(q-p_2)_\rho(q-p_2)_{\rho'}}{M_{D^*}^{2}}}{(q-p_2)^2-M_{D^*}^{2}}\left[g^{\rho\nu'}(q\cdot p_2)\right]\\
&\times \frac{g_{\nu'\nu}-\frac{q_\rho q_{\rho'}}{M_{D_1}^{2}}}{q^2-M_{D_1}^{2}}g^{\mu\nu}\frac{1}{(q-p_1)^2-M_{D}^{2}}\ \mbox{. }
\end{split}
\end{equation}

{We should mention that in Eqs.~\ref{DDstartriampl} and \ref{Jpitriampl} there are three-point tensor functions, which could be reduced to one point, two point and three point functions. The latter is convergent, while the former two can be divergent. Besides, each bubble in the bubble chains contains divergence as well. In this paper we use the dimensional regularization method with $\overline{\text{MS}}-1$ scheme to handle the divergences. The substraction constant of $\overline{\text{MS}}-1$ scheme is
\[
R=-\frac{1}{\epsilon}+\gamma_E-\ln(4\pi)-1\ \mbox{, }
\]
where $\epsilon=2-D/2$ with $D$ being the number of dimensions, and $\gamma_E$ is the Euler constant. A renormalization scale $\mu$ is also indispensable. The expressions of Eqs.~\ref{DDstartriampl} and \ref{Jpitriampl} after Passarino-Veltman reduction are too complicated to be shown here, but the $R$ and $\mu$ dependence of general one and two-point functions are shown as follows: for one-point functions
\begin{equation}
\begin{split}
A_0(M_a^2)&=\frac{\mu^{2\epsilon}}{i}\int\frac{d^Dk}{(2\pi)^D}\frac{1}{k^2-M_a^2+i0^+}\\
&=-\frac{M_a^2}{16\pi^2}\left(R+\ln\frac{M_a^2}{\mu^2}\right)\ \mbox{; }
\end{split}
\end{equation}
and for two-point functions
\begin{equation}
\begin{split}
&B_0(p^2,M_a^2,M_b^2)=\frac{\mu^{2\epsilon}}{i}\int\frac{d^Dk}{(2\pi)^D}
\frac{1}{(k^2-M_a^2+i0^+)\big[(p-k)^2-M_b^2+i0^+\big]}\\
&=\frac{1}{16\pi^2}\bigg[-R+1-\ln\frac{M_a^2}{\mu^2}+\frac{M_a^2-M_b^2-p^2}{2p^2}\ln\frac{M_b^2}{M_a^2}
+\frac{p^2-(M_a-M_b)^2}{p^2}\alpha(p^2)\ln\frac{\alpha(p^2)-1}{\alpha(p^2)+1}\bigg]\ \mbox{, }
\end{split}
\end{equation}
with $\alpha(p^2)=\sqrt{\frac{p^2-(M_a+M_b)^2}{p^2-(M_a-M_b)^2}}$.
}

Lastly, the non-polarized modulus-square can be written as
\begin{equation}
\begin{split}
&|\overline{\mathcal{M}}|^2=-\frac{4\pi\alpha |c_1|^2g^{\alpha\beta}}{3p_1^2 | p_1^2-M_X^2+iM_X\Gamma(p_1^2)|^2}\\
&\times\Big(g_{\alpha\mu}-\frac{p_{1\alpha}p_{1\mu}}{M_X^2}\Big)T_{DD^*,J/\psi\pi}^{\mu\sigma}\Big(g_{\beta\nu}-\frac{p_{1\beta}p_{1\nu}}{M_X^2}\Big)
T_{DD^*,J/\psi\pi}^{\nu\rho*}\Big(g_{\sigma\rho}-\frac{p_{D^*,J/\psi\sigma}p_{D^*,J/\psi\rho}}{M_{J/\psi,D^*}^2}\Big)\ \mbox{. }
\end{split}
\end{equation}

\end{document}